\documentclass[aps,prd,superscriptaddress,amsfonts,amssymb,amsmath,eqsecnum,nofootinbib,twocolumn,floatfix]{revtex4-2}
  \usepackage{color,graphicx}
  \usepackage[utf8]{inputenc}
  \usepackage[T2A]{fontenc}
  \usepackage[english]{babel}
  \definecolor{darkblue}{rgb}{0,0,0.7}
 \definecolor{darkred}{rgb}{0.7,0,0}
 \definecolor{darkgreen}{rgb}{0,0.4,0}
 \usepackage[unicode, colorlinks, citecolor=darkblue, linkcolor=darkred, urlcolor=blue]{hyperref}

 \allowdisplaybreaks
\begin{document}

\author{Aleksandr A. Movsisian}

\affiliation{Faculty of Physics, M.V. Lomonosov Moscow State University, Leninskie Gory, Moscow 119991, Russia}

\author{Albert I. Nazmiev}

\affiliation{Faculty of Physics, M.V. Lomonosov Moscow State University, Leninskie Gory, Moscow 119991, Russia}

\author{Andrey B. Matsko}

\affiliation{Jet Propulsion Laboratory, California Institute of Technology, 4800 Oak Grove Drive, Pasadena, California 91109-8099, USA}

\author{Sergey P. Vyatchanin}
\affiliation{Faculty of Physics, M.V. Lomonosov Moscow State University, Leninskie Gory, Moscow 119991, Russia}
\affiliation{Quantum Technology Centre, M.V. Lomonosov Moscow State University, Leninskie Gory, Moscow 119991, Russia}
\affiliation{Faculty of Physics, Branch of M.V. Lomonosov Moscow State University in Baku,\\ 1 Universitet street, Baku, AZ1144, Azerbaijan}

\date{\today}
	
\title{Broadband Multidimensional Variational Measurement with Non-Symmetric Coupling}

\begin{abstract}
 A broadband multidimensional variational measurement allows  overcoming the Standard Quantum Limit (SQL) of a classical mechanical force detection for a mechanical oscillator. In this measurement quantum back action, which perturbs the evolution of a mechanical oscillator, can be completely removed in a broad detection frequency band after post-processing. The measurement is performed by optical pumping of the central optical mode and analyzing the light escaping the two other optical modes, which have the frequency separation with the central mode equal to the mechanical frequency. To realize such a scheme in practice one either needs to use a very long optical interferometer or should utilize optical modes belonging to different mode families. In the second case the modes have different geometries and their coupling with the mechanical mode is not identical. Here we analyze a general case of the non-symmetric measurement scheme, in which the coupling strengths with the light modes are not equal to each other, and take into account optical losses. We found that the back action can be completely excluded from the measurement result in the case of the asymmetric lossless system. The nonzero loss limits the sensitivity. An experimental implementation of the proposed scheme is discussed. 
\end{abstract}

\maketitle

\section{Introduction}

Optical transducers are used to characterize mechanical motion, which changes a quadrature amplitude of the probe light. The sensitivity of the measurement can be extremely high. For instance, it was demonstrated that optical gravitational wave detectors are capable of measuring the relative mechanical displacement much smaller than a proton size \cite{AbbotLRR2020,aLIGO2015,MartynovPRD16,AserneseCQG15, DooleyCQG16,AsoPRD13, AkutsuPTEP2021}. The search for better detection methods is still actual nowadays. In this paper we study theoretically a realistic optical detection scheme that can potentially improve sensitivity of the optical detection of a mechanical force. 

The optical detector of a classical force involves optical and mechanical subsystems. The  thermal fluctuations in the mechanical probe system (Nyquist noise) and quantum noise of the meter introduce fundamental limits of the measurement sensitivity. The impact of the thermal noise can be considerably reduced if one measures a variation of the position during a time period that is much smaller than the mechanical ring down time \cite{Braginsky68, 92BookBrKh}. However, one cannot avoid the thermal noise that is spectrally overlapping with the signal and coming to the system from the same channel as the signal. The quantum optical noise restricts sensitivity by the well-known standard quantum limit (SQL) \cite{Braginsky68, 92BookBrKh}. SQL is a consequence of noncommutativity between the optical probe noise and the quantum back action noise associated with the ponderomotive action of the probe light on the mechanical system. However there are several ways to overcome it. 

In this work we study the optimization of the measurement technique involving a few optical frequency harmonics. This method allows beating SQL. For instance, a dichromatic optical probe may lead to observation of such phenomena as negative radiation pressure \cite{Povinelli05ol,maslov13pra} and optical quadrature-dependent quantum back action evasion \cite{21a1VyNaMaPRA}. 

A multidimensional variational measurement \cite{22PRAVyNaMa} is one of methods that involve a few optical harmonics. Advantage of the measurement is its relatively broad bandwidth. In such a measurement the force of interest acts on a mechanical oscillator. The mechanical oscillator is coupled to a system with three optical modes,  whose frequencies $\omega_\pm,\ \omega_0$ are separated by the mechanical frequency $\omega_m$ so that $\omega_\pm =\omega_0\pm \omega_m$. The measurement is performed by optical pumping of the central optical mode $\omega_0$ and measuring the light escaping the two other modes $\omega_\pm$. Detection of optimal quadrature components of the output waves of modes $\omega_\pm$ {\em separately} provides two channels registration. It allows to detect back action and to remove it {\em completely in broad band} from the measured data via post processing. 

Previous studies of the multidimensional variational measurements involved exclusively ideal and symmetrical systems. However, practical realization of the idea calls for usage of a not ideal configuration. To prove practical validity of the measurement scheme we here further investigate the sensitivity of broadband multidimensional variational measurement in the presence of optical losses. We also consider the non-symmetric case when the relaxation rates as well as couplings of the optical modes $\omega_\pm$ with probe mechanical oscillator are not equal to each other. We found that it is still possible to suppress the back action in the case of zero loss and the asymmetry present. The loss limits the sensitivity, but the SQL still can be beaten. The result proves robustness of the measurement technique. We also discuss the feasibility of separation of the optical modes of interest escaping the mechanical resonator. This particular problem is hard because the reasonable frequency separation between the modes in the optical triplet is less than a MHz.

\section{Physical Model}
\label{Model}

\begin{figure}
\includegraphics[width=0.45\textwidth]{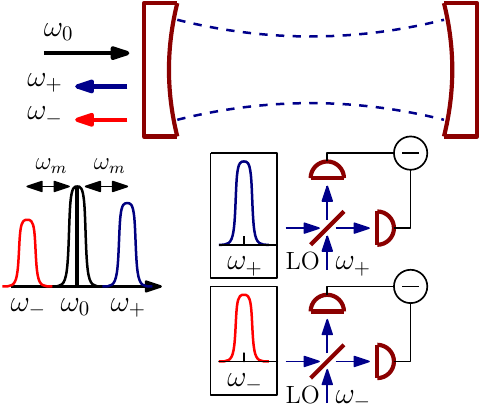}
\caption{Frequencies $\omega_0,\ \omega_\pm$ of three optical modes in the cavity are separated by the frequency $\omega_m$ of mechanical oscillator. Optical modes are coupled to the mechanical oscillator via optical pressure. The relaxation rates of the optical modes are smaller than mechanical frequency $\gamma,\ \gamma_\pm\ll \omega_m$. The middle mode with frequency $\omega_0$ is resonantly pumped. The outputs of modes $\omega_\pm$ are detected separately.}\label{schemeM}
\end{figure}

Let us consider an optical cavity with a triplet of optical modes characterized with resonant frequencies $\omega_-,\ \omega_0,\ \omega_+$, separated by eigen frequency $\omega_m$ of mechanical oscillator (Fig.~\ref{schemeM}). We assume that the relaxation rates of the optical modes are different and characterized by the full width at the half maxima (FWHM) equal to 
\begin{subequations}
	\label{gammaDef}
\begin{align}
2\gamma_\pm &=2(\gamma_{0\pm}+\gamma_{e\pm}),\\
2\gamma &= 2\gamma_0+\gamma_e,
\end{align}
\end{subequations}
for modes $\omega_\pm$ and for $\omega_0$, respectively. 
Here $\gamma_{0\pm},\ \gamma_0$ characterize effective transmittance of the input mirror and $\gamma_e,\ \gamma_{e\pm}$ stand for optical losses of the cavity. We assume that the relaxation rates $\gamma_e,\ \gamma_{e\pm}$ are small if compared with the transmission ones
\begin{align}
\label{Smallgammae}
    \gamma_e,\ \gamma_{e\pm} \ll \gamma_0,\ \gamma_{0\pm}.
\end{align}
The mechanical relaxation rate $\gamma_m$ is small in comparison with the optical one. We also assume that the optical triplet is symmetric and matched with the mechanical oscillator. The conditions of the resolved side band interaction and the condition of optical loss smallness are also valid.
\begin{align}
	\label{RSB}
	\omega_\pm =\omega_0\pm \omega_m, \quad  \gamma_m\ll \gamma \ll \omega_m. \quad 
\end{align}

There are two external actions on the opto-mechanical system. The central mode with frequency $\omega_0$ is resonantly pumped and a classical force of interest is acting on the mechanical mass. The optical modes  $\omega_\pm$ are not pumped. Mass of mechanical oscillator is a movable end mirror, which provides coupling with optical modes.  We detect the output of optical modes $\omega_\pm$ to measure the force (Fig.~\ref{schemeM}). The quadrature components of the sideband modes are measured separately by balanced homodyne detectors with corresponding local oscillators having frequencies coincident with the frequencies of the sideband modes $\omega_\pm$. The signal is inferred by post-processing of the linear combination of the measured results.

The generalized Hamiltonian describing the system can be presented in form
\begin{subequations}
   \label{Halt}
  \begin{align}
  H   &= H_0 + H_\text{int}+H_s + \\
  & +H_{T 0}+H_{\gamma_0} + H_{T e}+H_{\gamma_e}+H_{T, \, m}+H_{\gamma_ m} ,\nonumber\\
  \label{H0}
  H_0 &=\hslash \omega_+\hat c_+^\dag \hat c_+  + \hslash \omega_0 \hat c_0^\dag \hat c_0 +\\
    &\qquad + \hslash \omega_-\hat c_-^\dag \hat c_-
         +\hslash \omega_m \hat d^\dag \hat d,\nonumber\\
  \label{Hint}
    H_\text{int} & = \frac{\hslash }{i}
      \left( \left[\eta_-\hat c_0^\dag \hat c_- + \eta_+\hat c_+^\dag \hat c_0\right] \hat d -\right.\\
     &\qquad -\left.  \left[\eta^*_-\hat c_0 \hat c_-^\dag+ \eta^*_+\hat c_+ \hat c_0^\dag \right]\hat  d^\dag \right),\nonumber\\
   \label{Hs}\hat 
   H_s & = - F_s x_0\left(\hat d+ \hat d^\dag\right). 
  \end{align}
  \end{subequations}
Here $\hslash$ is Plank constant. $H_0$ describes energies of the optical modes and the mechanical oscillator, $\hat c_0,\ \hat c^\dag_0,\ \hat c_\pm,\ \hat c^\dag_\pm$ are annihilation and creation operators for the corresponding optical modes, $\hat d,\ \hat d^\dag$ are annihilation and creation operators of the mechanical oscillator. The operator of coordinate $\hat x$ of the mechanical oscillator is presented in form
\begin{align}
\label{x}
 \hat x= x_0\left(\hat d + \hat d^\dag\right),\quad x_0=\sqrt\frac{\hslash}{2m \omega_m},
\end{align}
where $m$ is the mass of the oscillator. $H_\text{int}$ is the interaction Hamiltonian written in the rotation wave approximation for the optical and mechanical modes. \footnote{
The canonical interaction Hamiltonian $H_\text{int}\sim (E_0+E_+ +E_-)^2x$, where $E_0,E_-,E_+$ are electric fields of modes $0,-,+$ on surface of mirror (mass of oscillator), can be transformed into form \eqref{Hint} after omitting fast oscillating terms.} We here consider non symmetric interaction and introduce different coupling constants $\eta_\pm\simeq x_0\omega_0/L$, where  $L$ is the length of the cavity. $H_s$ is a part of Hamiltonian describing  signal force $F_s$.   
$H_{T0}$ is the Hamiltonian describing the outer environment (regular and fluctuational fields incident on the input mirror) and  $H_{\gamma_0}$ is the Hamiltonian of the coupling between the outer environment and the optical modes, resulting in decay rate $\gamma_0$; $H_{Te}$ and $H_{\gamma_e}$ describe optical losses. The pump is also included into $H_{\gamma_0}$. Similarly, $H_{T, \, m}$ is the thermal bath Hamiltonian and $H_{\gamma_m}$ is the Hamiltonian describing coupling between the environment and the mechanical oscillator resulting in a decay rate $\gamma_m$. 

\section{Analysis }

\subsection{Basic equations}

We denote the values of the input and output optical amplitudes as $\hat a_{\pm, \,0}$ and  $\hat b_{\pm, \,0}$, respectively. Using the Hamiltonian \eqref{Halt} we derive the equations of motion for the intracavity slow amplitudes of fields.
\begin{subequations}
\begin{align} 
	\label{c0}
\dot {\hat c}_0+\gamma \hat c_0&=\eta_+^*\hat c_+ \hat d^\dag - \eta_- \hat c_- \hat d+\\
 &\qquad  +\sqrt{2 \gamma_0}\,\hat a_0 + \sqrt{2 \gamma_e}\, \hat e_0,\nonumber\\
\dot {\hat c}_++\gamma_+ \hat c_+&=-\eta_+ \hat c_0 \hat d 
	+\sqrt{2 \gamma_{0+}}\,\hat a_+ +\sqrt{2\gamma_{e+}} \hat e_+, \\
\dot { \hat c}_-+\gamma_- \hat c_-&=\eta_-^*\hat c_0 \hat d^\dag 
	 + \sqrt{2 \gamma_{0-}}\,\hat a_-+\sqrt{2 \gamma_{e-}}\,\hat e_-,\nonumber\\
\dot {\hat d}+\gamma_m \hat d&=\eta^*_-\hat c_0 \hat c_-^\dag + \eta^*_+\hat c_0^\dag \hat c_+  +\sqrt{2 \gamma_m}\,\hat q +f_s.
\end{align}  \label{moveq}
\end{subequations} 
Here operators $\hat e_\pm$ describe quantum fluctuations due to optical losses, see definitions \eqref{gammaDef},
$\hat q$ is the fluctuation force acting on mechanical oscillator, and $f_s$ is the signal force. The temporal structure of the force is defined in what follows.  

The operators $\hat a_\pm,\ \hat e_\pm,\ \hat q$ are characterized by the following commutators and correlators
\begin{subequations}
 \label{commT}	
\begin{align}
	 \label{commA}
	\left[\hat a_\pm(t), \hat a_\pm^\dag(t')\right] &=
	\left\langle\hat a_\pm(t)\, \hat a_\pm^\dag(t')\right\rangle = \delta(t-t'),\\
	\label{commE}
	\left[\hat e_\pm(t), \hat e_\pm^\dag(t')\right] &=  
	\left\langle\hat e_\pm(t)\, \hat e_\pm^\dag(t')\right\rangle = \delta(t-t'),\\
	\left[\hat q(t), \hat q^\dag(t')\right] &=   \delta(t-t'),\\
	\label{corrQ}
	\left\langle\hat q(t) \hat q^\dag(t')\right\rangle &= (2n_T +1)\, \delta(t-t'),\\
	& n_T= \left ( e^{\hslash \omega_m/\kappa_BT} -1 \right )^{-1}.
\end{align}
\end{subequations}
Here $\langle \dots \rangle$ stands for the ensemble averaging.  $n_T$ is the thermal number of mechanical quanta, $\kappa_B$ is the Boltzmann constant,and $T$ is the ambient temperature.

The input-output relations connecting the incident ($\hat a_\pm$) and intracavity ($\hat c_\pm$) amplitudes with output ($\hat b_\pm$) amplitudes are
\begin{align}
 \label{outputT}
  \hat b_\pm= -\hat a_\pm + \sqrt{2\gamma_{0\pm}} \hat c_\pm.
\end{align} 

It is convenient to separate the expectation values of the wave amplitudes at frequency $\omega_0$ (described by block letters) as well as its fluctuation part (described by small letters) and assume that the fluctuations are small:
 \begin{align}
 \label{expA}
 \hat c_0 & \Rightarrow C_0 +  c_0 ,\quad  
 \end{align}
here $C_0$  stands for the expectation value of the field amplitude in the central optical mode and $c_{  0}$ represent the quantum fluctuations of the field confined in the mode, $|C_0|^2 \gg \langle  c_0^\dag  c_0 \rangle$. Similar expressions can be written for the sideband optical modes as well as the mechanical mode. The normalization of the amplitudes is selected so that $\hslash \omega_0 |A_0|^2$ describes the optical power of incident wave \cite{02a1KiLeMaThVyPRD}. 

We assume in what follows that the expectation amplitudes are real, same as the coupling constants $\eta_\pm$
\begin{align}
	\label{real}
	A_0=A_0^*,\quad C_0= C_0^*=\sqrt{\frac{2}{\gamma_0}}A_0,\quad  \eta_\pm=\eta_\pm^*.
\end{align}

The Fourier transform of  operators, for example, $\hat a_\pm$ is defined as follows
\begin{subequations}
\begin{align}
 \label{apmFT}
 \hat a_\pm (t) &= \int_{-\infty}^\infty a_\pm(\Omega) \, e^{-i\Omega t}\, \frac{d\Omega}{2\pi}.
\end{align}
For operators $a_\pm(\Omega)$ the following commutators and correlators are valid:
\begin{align}
 \label{comm1}
  \left[a_\pm(\Omega), a_\pm^\dag(\Omega')\right] &= 2\pi\, \delta(\Omega-\Omega'),\\
  \label{corr1}
  \left\langle a_\pm(\Omega) a_\pm^\dag(\Omega')\right\rangle &= 2\pi\, \delta(\Omega-\Omega')
\end{align}
\end{subequations}
Similar expressions can be written for the other noise operators ($\hat e_\pm,\ \hat q$). 

Substituting \eqref{expA} and \eqref{real} into the equations of motion \eqref{moveq} and keeping only terms of first order of smallness, we obtain 
\begin{subequations}
	\label{set1}
	\begin{align}
		(\gamma_+ -i\Omega)c_+(\Omega)&=-\eta_+ C_0  d(\Omega) +\\
		& +\sqrt{2 \gamma_{0+}}\, a_+ (\Omega) +\sqrt{2\gamma_{e+}}  e_+(\Omega), \nonumber\\
		(\gamma_- -i\Omega) c_-(\Omega)&=\eta_+ C_0  d^\dag(-\Omega)  +\\
		& + \sqrt{2 \gamma_{0-}}\, a_-(\Omega)+\sqrt{2 \gamma_{e-}}\, e_-(\Omega),\nonumber\\
		(\gamma_m -i\Omega) d(\Omega)&= C_0\big[\eta_-  c_-^\dag(-\Omega) +  \eta_+ c_+(\Omega)\big]+\\
		&\qquad  +\sqrt{2 \gamma_m}\,\hat q(\Omega) +f_s(\Omega).\nonumber
	\end{align}
\end{subequations}
Outputs $b_\pm$ around frequencies $\omega_\pm$ have to be detected separately. We also see that fluctuation waves around $\omega_0$ do not influence on field components in the vicinity of frequencies $\omega_\pm$ and the first equation \eqref{c0} does not couple with others, so we omit it in the further consideration. 

We assume that the signal force is a resonant square pulse acting during time interval $\tau$ ($\omega_m\tau\gg 1$):
\begin{align}
\label{Fs}
F_S(t)&= F_{s0}\sin(\omega_m t + \psi_f) = \\
    = & i \left(F_{s}(t) e^{-i\omega_m t} - F_{s}^*(t) e^{i\omega_m t}\right),\quad 
  -\frac{\tau}{2} < t < \frac{\tau}{2},\nonumber\\
  \label{fs}
  f_s(\Omega)& = \frac{F_s(\Omega)}{\sqrt{2\hslash \omega_m m}},\\
  f_{s0}(\Omega)& = \frac{F_{s0}(\Omega)}{\sqrt{2\hslash \omega_m m}}= 2f_s(\Omega).
\end{align}
where  $F_s(\Omega)\ne F_s^*(-\Omega)$ is the Fourier amplitude of $F_s(t)$.  

Let us introduce quadrature amplitudes of amplitude and phase
\begin{subequations}
\label{quadDef}
 \begin{align}
  a_{\pm a} &= \frac{a_\pm (\Omega) +a_\pm ^\dag(-\Omega)}{\sqrt 2}\,,\\
	 a_{\pm \phi} &= \frac{a_\pm (\Omega) -a_\pm ^\dag(-\Omega)}{i\sqrt 2}\,.
 \end{align}
\end{subequations}
Using \eqref{set1} we obtain equations for amplitude quadratures
\begin{subequations}
	\label{quadAmp}
	\begin{align}
		\label{ca+DDNS}   
		(\gamma_+ - i\Omega)c_{+a} &+  \eta_+ C_0  d_a = \sqrt {2 \gamma_{0+}}  { a}_{ +a} +\\
		& + \sqrt{2\gamma_{e+}} e_{+a},\nonumber\\
		\label{ca-DDNS} 
		(\gamma_- - i\Omega) c_{-a} &- \eta_- C_0  d_a = \sqrt {2 \gamma_-}  { a}_{-a}+\\
		& + \sqrt{2\gamma_{e-}} e_{-a},\nonumber\\
		\label{daDDNS}
		(\gamma_m - i\Omega)  d_a &-  C_0 \Big(\eta_+c_{+a}+ \eta_-c_{-a}\Big)=\\
		&= \sqrt {2 \gamma_m} q_a +  f_{s\,a},\nonumber
	\end{align}
\end{subequations}
and for phase quadratures
\begin{subequations}
	\label{quadPhase}
   \begin{align}
		\label{cphi+DDNS}   
		(\gamma_+ - i\Omega)c_{+\phi} &+  \eta_+ C_0  d_\phi = \sqrt {2 \gamma_{0+}}  { a}_{ +\phi}+\\
		& + \sqrt{2\gamma_{e+}} e_{+\phi},\nonumber\\
		\label{cphi-DDNS} 
		(\gamma_- - i\Omega) c_{-\phi} &+ \eta_-  C_0  d_\phi = \sqrt {2 \gamma_-}  { a}_{-\phi}+\\
			& + \sqrt{2\gamma_{e-}} e_{-\phi},\nonumber\\
		\label{dphiDDNS}
		(\gamma_m - i\Omega)  d_\phi & -  C_0 \Big(\eta_+c_{+\phi} - \eta_- c_{-\phi}\Big)= \\
	    &= \sqrt {2 \gamma_m} q_{\phi} +  f_{s\,\phi}.\nonumber
	\end{align}
\end{subequations}

Set \eqref{quadAmp} for amplitude quadratures does not depend on set \eqref{quadPhase} for phase quadratures, so we analyze the set for amplitude quadratures only.

\subsection{Quantum noise spectral density}

Detailed derivation of the expressions for the output quadratures is presented in Appendix \ref{appPSD}. Here we analyze the resultant expressions and derive expression for the spectral density of the quantum noise of the meter. 

Inequality of $\eta_\pm$ and $\gamma_\pm$ introduces classical dynamic back action manifesting by the optical damping rate $G$
\begin{align}
G  = \left[\frac{\eta_+^2}{\gamma_+-i\Omega}- \frac{\eta_-^2}{\gamma_- -i\Omega}\right] C_0^2 .\label{GDNS}
\end{align}
The complete mechanical relaxation rate $\Gamma_m=\gamma_m+G$ is a sum of the intrinsic relaxation rate $\gamma_m$ and the introduced optical damping $G$.

Assuming that we are able to measure $b_{+a}$ and $b_{-a}$ separately (the quadratures are not commuting Hermitian operators), present their weighted sum $\Sigma$ as
\begin{align}
  \label{Sigma1}
  \Sigma = z_+ \, \frac{b_{+a}\big(\Gamma_m -i\Omega\big)}{A_+\big(z_- -z_+\big)}+ z_- \, \frac{b_{-a}\big(\Gamma_m -i\Omega\big)}{A_-\big(z_- -z_+\big)},
\end{align}
introducing coefficients 
\begin{subequations}
 \begin{align}
   \label{Bpm}
   A_\pm &= \frac{\sqrt{2\gamma_{0\pm}}\,\eta_\pm C_0}{(\gamma_\pm -i\Omega)}, \ B_{e\pm}=\frac{\sqrt{2\gamma_{0\pm}}(\Gamma_m-i\Omega)}{\eta_\pm C_0},\\
      B_\pm &=\frac{(\gamma_{0\pm} -\gamma_{e\pm} +i\Omega)(\Gamma_m-i\Omega)}{\sqrt{2\gamma_{0\pm}}\,\eta_\pm C_0},\\
		\label{Ypm}
		&Y_\pm=\frac{A_\pm}{B_\pm} ,\quad Y_{e\pm}= \frac{A_\pm}{B_{e\pm}}
\end{align}
\end{subequations}

The complex functions $z_\pm (\Omega)$, depending on spectral frequency $\Omega$,  are free post-processing parameters, which should be selected in an optimal way. Due to homogeneity of the weighted sum $\Sigma(\lambda z_+, \lambda z_-) = \lambda \Sigma(z_+,z_-)$, we can introduce one parameter $y$ instead of two parameters $z_\pm$ in the optimization task,
\begin{equation}
y+\frac 1 2= \frac{z_-}{z_--z_+} ,\quad y-\frac 1 2  =\frac{z_+}{z_--z_+}.
\end{equation}

Assuming that the input fields $a_\pm$ are in the vacuum states and using \eqref{Sigma1}, we write for the quantum noise spectral density:
\begin{subequations}
	\label{SfLosses}
	\begin{align}
		S_f & = S_{qu} + S_T,\\
		S_{qu} &=|B_+|^2\left|y-\frac 1 2 + Y_+\right|^2 
		+ |B_-|^2\left|y+\frac 1 2 + Y_-\right|^2 +\nonumber\\
		\label{Bep}
		&+\frac{\gamma_{e+}}{\gamma_{0+}}|B_{e+}|^2\left|y-\frac 1 2 + Y_{e+}\right|^2+\\
		\label{Bem}
		& + \frac{\gamma_{e-}}{\gamma_{0-}}|B_{e-}|^2\left|y+\frac 1 2 +Y_{e-}\right|^2 ,\\
		S_T &= 2\gamma_m \left(n_T+ \frac 1 2\right).
	\end{align}
\end{subequations}
Here  $S_{qu}$ relates to quantum noise, $S_T$ stands for the thermal noise in \eqref{SfLosses}, the terms \eqref{Bep}, \eqref{Bem} appear due to the optical losses. The first two terms in $S_{qu}$ describe main optical noise,  values $Y_\pm$ are generated due to back action. 

Let us assume that the condition \cite{Braginsky68, 92BookBrKh} of smallness of thermal noise compared with SQL
\begin{align}
	\label{BrCond}
	B= \frac{n_T \omega_m\tau}{Q}\ll 1
\end{align}
is fulfilled ($Q$ is a mechanical quality factor). The main requirement for this is a large ring down time $\gamma_m^{-1}$ and a fast interrogation time $\tau$, i.e. $\gamma_m\tau\ll 1$. For the parameters listed in Table~\ref{table1} factor $B\simeq 0.2$.

\subsection{Symmetric system}

In the simplest case of lossless ($\gamma_{e\pm}=0$, terms \eqref{Bep}, \eqref{Bem} disappear) and symmetric  ($B_+=B_-\equiv B, \ Y_+=Y_-\equiv Y$) system the back action can be excluded {\em completely} by optimally choosing $y_\text{opt}=-Y$:
\begin{align}
	\label{symNoLoss}
	S_{qu}|_\text{noLoss}^\text{sym} &= |B|^2/2 = \frac{\gamma_m^2 +\Omega^2}{\mathcal K}, \quad y_\text{opt} = -Y,\\
	& \mathcal K=\frac{4\gamma\eta^2C_0^2}{\gamma^2+\Omega^2}.
\end{align}
In this case back action is completely removed and spectral density monotonically decreases with power ($\sim \mathcal K$) and can be made smaller $S_{SQL}=2\sqrt{\gamma_m^2+\Omega^2}$ \eqref{SSQL} \cite{21a1VyNaMaPRA} (see reminder on SQL in Appendix \ref{appSQL}).

\subsection{Measurement optimization}

In general case in order to find the minimum of $S_{qu}$ we introduce the optimal complex function $y_\text{opt}$
\begin{subequations}
	\begin{align}
	  y_\text{opt} & =  \frac{\left| B_+\right|^2}{\mathbb B^2}\left(\frac 1 2  - Y_+\right) 
		-\frac{\left| B_-\right|^2}{\mathbb B^2}\left(\frac 1 2  + Y_-\right)
		+ \Delta_e ,\\
		\label{mathbbB}
	\mathbb B^2 &= \left| B_+\right|^2 +\left| B_-\right|^2 + \mathbb B_e^2,\\ \mathbb B_e^2 &= \frac{\gamma_{e+}}{\gamma_{0+}}|B_{e+}|^2 + \frac{\gamma_{e-}}{\gamma_{0-}}|B_{e-}|^2 \,,\\
		\label{Deltae}
	\Delta_e &= \left[\frac{\gamma_{e+}}{\gamma_{0+}}\right] \frac{\left| B_{e+}\right|^2}{\mathbb B^2}\left(\frac 1 2  - Y_{e+}\right) -\\
		&\quad 	- \left[\frac{\gamma_{e-}}{\gamma_{0-}}\right] \frac{\left| B_{e-}\right|^2}{\mathbb B^2}\left(\frac 1 2  + Y_{e-}\right)\, .
	\end{align} 	
\end{subequations}
It should be substituted into \eqref{SfLosses} to find the minimum achievable noise. In what follows we consider a few realizations of the measurement systems and find the optimal measurement sensitivity for those cases. 

\begin{figure}
\includegraphics[width=0.48\textwidth]{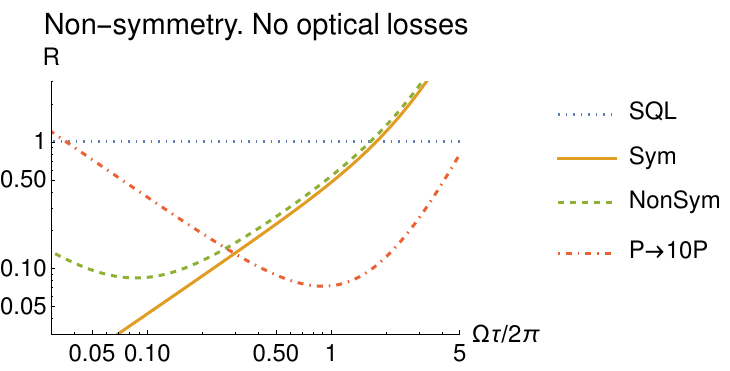}
	\caption{Plots of ratio $R = S_{qu}/S_{SQL}$, of spectral densities {\em without} optical losses for symmetric \eqref{noLoss} (with $\gamma_+=\gamma_-, \ \eta_+=\eta_-$) and non-symmetric cases  \eqref{noLoss} as a function of spectral frequency $\Omega$,
	$\tau$ is the time of signal force action, other parameters are taken from Table~\ref{table1}. Plot marked as $P\to 10P$ corresponds to the non-symmetric case with 10 times increased pump $P_{in}$.}\label{NonSymNoLoss}
\end{figure}

\begin{table}[b]
	\caption{Parameters of a mechanical oscillator (SiN membrane) and optical cavity, used for estimates.} \label{table1}
	\begin{tabular}{||c | c | c||}
		\hline
		\multicolumn{3}{||c||}{Membrane} \\
		\hline
		Mass, $m$ & 50 & $10^{-9}$ g\\
		Frequency, $\omega_m/2\pi$& 350 & $10^3$ Hz\\
		Quality factor $Q=\frac{\omega_m}{2\gamma_m}$ & $10^9$ & \\
		Temperature, $T$ & 20 & K$^\circ$ \\
		Thermal phonons number, $n_T$ & $1.2\cdot 10^6$ &\\
		Time of signal force $\tau=30 \cdot\frac{2\pi}{\omega_m}$ & $0.84\cdot 10^{-3}$ & sec \\
		\hline
		\multicolumn{3}{||c||}{Cavity} \\
		\hline
		Length of cavity & 10 & cm \\
		Bandwidth , $\gamma_0 + \gamma_e$ & 2.3 $10^{5}$ & s$^{-1}$\\
		$\gamma_e,\ \gamma_{e\pm}$&  2.3 $10^{3}$ & s$^{-1}$\\
		$\gamma_\pm$&   $(1\mp 1\%)\gamma_0$ & \\
		Couplings $\eta_\pm$ & $\frac{\omega_0x_0}{L}(1\pm 3\%)$ & \\
		Wave length, $\lambda= 2\pi c/\omega_0$ & 1.55 & $10^{-6}$ m \\ 
		Input power $P_{in}$ & 1 & $10^{-6}$ W \\
		\hline
	\end{tabular}
\end{table}

\subsubsection{Non-symmetric case without optical losses}

For the non-symmetric case ($\eta_+\ne \eta_-$) and for the absence of optical losses ($\gamma_{e\pm}=0$) the formulas are more compact
\begin{align}
	\label{noLoss}
	S_{qu}|_\text{noLoss}^\text{non-sym} &= \frac{\left| B_+\right|^2\left| B_-\right|^2 }{\left| B_+\right|^2+\left| B_-\right|^2} \left|1 - Y_+ + Y_-\right|^2,\\
  y_\text{opt}=  \frac{\left| B_+\right|^2}{\mathfrak B^2}&\left(\frac 1 2  - Y_+\right) 
  -\frac{\left| B_-\right|^2}{\mathfrak B^2}\left(\frac 1 2  + Y_-\right),\\
  \text{Here}\quad &\mathfrak B^2= \left| B_+\right|^2+\left| B_-\right|^2
\end{align} 
We see that in non-symmetric case back action can be excluded only partially (term $(Y_+-Y_-)$ in \eqref{noLoss}). 

In nearly-resonant case, $\gamma_\pm\gg \Omega$, formula \eqref{noLoss} can be simplified using definitions \eqref{GDNS} and \eqref{Bpm}
\begin{align}
	\label{noLoss2}
	S_{qu}|_\text{noLoss}^\text{non-sym} &\underbrace{\simeq}_{\gamma_\pm \gg \Omega} \frac{1 }{2 G_+} \left(\big( \gamma_m - G)^2+\Omega^2 \right),\\
	 G_+= \frac{\eta_+^2C_0^2}{\gamma_+} & + \frac{\eta_-^2C_0^2}{\gamma_-}, \quad
	 G\simeq \frac{\eta_+^2C_0^2}{\gamma_+} - \frac{\eta_-^2C_0^2}{\gamma_-}.
\end{align} 

We see that the introduced optical damping $G$ can be zero even in the non-symmetric case, if $\eta_+^2/\gamma_+=\eta_-^2/\gamma_-$. The spectral density practically coincides with \eqref{symNoLoss}. It means that {\em in the lossless asymmetric measurement system the back action can be completely excluded} (at corresponding choice of $Y_\pm$). The normalized spectral density $S_{qu}$ monotonically decreases with the optical power increase. This is an important finding of our research.

A non-zero value $G$ of optical damping characterizes back action, which partially degrades sensitivity. For condition $\gamma_m \ll G,\ \Omega$ we have
\begin{align}
	\label{noLoss3}
	S_{qu}|_\text{noLoss}^\text{non-sym} &= \frac{G^2+\Omega^2}{2 G_+}\ge  \left|\frac{G\Omega }{ G_+} \right| < S_{SQL}\simeq 2|\Omega|.
\end{align}
So while in the non-symmetric case one can surpass SQL, the sensitivity becomes limited.

\begin{figure}
	\includegraphics[width=0.48\textwidth]{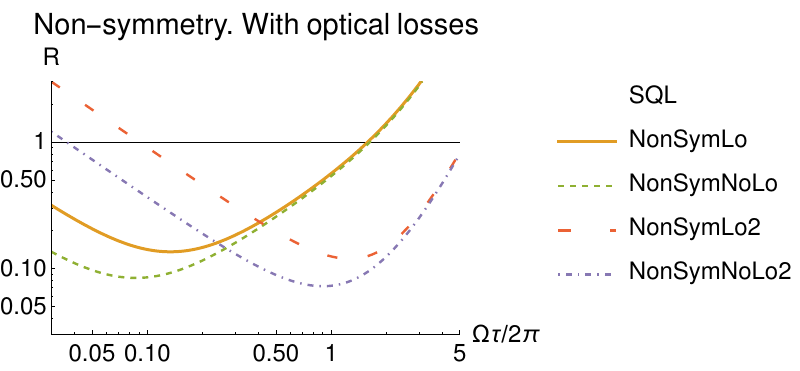}
	\caption{Plots of ratio $R = S_{qu}/S_{SQL}$, of spectral densities {\em with} optical losses for non-symmetric  cases  \eqref{SfLosses} as a function of spectral frequency $\Omega$. For the plot marked by "NonSymLo" the parameters are taken from Table~\ref{table1} (the dashed line "NonSymNoLo" correspond to same parameters but zero optical losses). Plots "NonSymLo2" and "NonSymNoLo2" corresponds to 10 times larger input pump $P_{in}$. $\tau$ is the time of signal force action.} \label{NonSymLoss2}
\end{figure}

\subsubsection{Non-symmetric case with optical losses}

We studied the most general case of the sensor will loss numerically. using the expressions \eqref{SfLosses} and considering the optimal function $y_{\text{opt}}$ \eqref{mathbbB}.
The plots of normalized spectral densities are presented in Fig.~\ref{NonSymLoss2} for the non-symmetric case with losses. The numerical parameters listed in Table~\ref{table1} were utilized. The dashed lines correspond to the same parameters but zero loss. 

Naturally, the sensitivity degrades rather significantly because of the loss present. However, for the 1\% attenuation in the cavity the observed sensitivity improvement is still significant. 

\begin{figure}
	\includegraphics[width=0.48\textwidth]{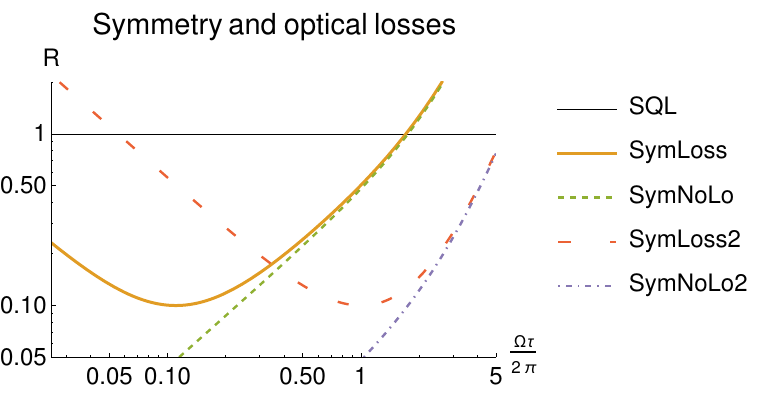}
	\caption{Plots of ratio $R = S_{qu}/S_{SQL}$, of spectral densities for the symmetric  case {\em with} optical losses. For plot marked by "SymLoss" the parameters are taken from Table~\ref{table1} (the dashed line "SymNoLo" corresponds to the same parameters but zero optical losses). Plots "SymLoss2" and "SymNoLo2" correspond to 10 times larger input pump $P_{in}$. $\tau$ is the time of signal force action.}\label{SymLoss}
\end{figure}

\subsubsection{Symmetric case with optical losses}

Optical losses restrict sensitivity even in the symmetric case. Corresponding plots are presented in Fig~\ref{SymLoss}. We see that even small losses (about only 1\% from losses through the input mirror) limit the sensitivity. The reason is the uncorrelated back action resulting from the additional noise due to optical losses which can not be completely subtracted.

\begin{figure}
	\includegraphics[width=0.48\textwidth]{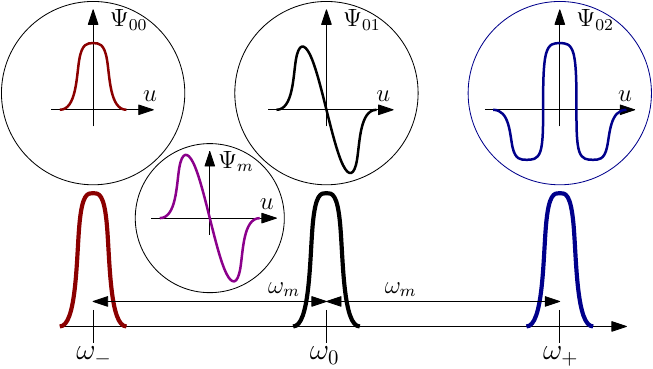}
	\caption{Higher order modes in a Fabry-Perot cavity can be used as modes $\omega_0,\ \omega_\pm$, their distribution functions $\Psi_{00}(u),\ \Psi_{01}(u),\ \Psi_{02}(u)$) are shown in the top row, $u$ is one of the transverse coordinates.}\label{SchemeBVM}
\end{figure}

\section{Physical realization}

There are a few practical problems with experimental realization of the proposed measurement scheme.
Those include i) difficulty to create a triplet with equidistant separation of modes $\omega_\pm$ from $\omega_0$; ii) achieve the value of separation $|\omega_0 -\omega_\pm|$ to be equal to the mechanical frequency $\omega_m$ about 1\dots100 MHz; iii) separate the optical harmonics without loss. It is rather a small frequency difference in the optical domain. Realization of such a narrow and lossless optical filter is a difficult experimental task. 

We can use optical modes belonging to different geometrical mode families and corresponding geometrical mode sorters \cite{Gu2018PRL, CarpenterNaureC2019} to realize the proposed system. 

Let us consider higher order modes (HOM) in a Fabry-Perot cavity. The frequency range between these modes can be made much less than the free spectral range (FSR) of the cavity by properly selecting the radii of curvature of mirrors. Fig.~\ref{SchemeBVM} illustrates such a possibility. Distribution functions of the modes $\Psi_{00}(u),\ \Psi_{01}(u),\ \Psi_{02}(u)$) are shown in the top row, where $u$ is one of the transverse coordinates, distributions over the other transverse coordinate $v$ are assumed to be Gaussian. 

These optical modes can interact with the dipole elastic mode of the mirror (membrane). Such a mechanical mode has geometrical distribution $\Psi_m$ illustrated by Fig.~\ref{SchemeBVM}. The generalized force acting on the elastic mode is defined by the overlap integral. In other words, the coupling constants $\eta_\pm$ is proportional to
\begin{align}
 \eta_+ &\sim \int \Psi_{01}(u)\,\Psi_{02}(u)\, \Psi_m(u)\, du,\\
 \eta_- &\sim \int \Psi_{01}(u)\,\Psi_{00}(u)\, \Psi_m(u)\, du
\end{align}
While the overlap integral is not zero, the absolute values of $\eta_+$ and $\eta_-$ can be different. As we have shown above, this asymmetry does not prevent the BAE measurement. 

Next step is to separate the modes leaving the cavity in space. The output light consists of radiation from all three optical modes. Theoretically, a mode sorter can do it perfectly, and experimental efficiency exceeding 80 \% was recently confirmed \cite{Gu2018PRL}. Therefore, the modes can be separated, reshaped to Gaussian beams and optimally detected to realize the proposed measurement.  

\section{Conclusion}

We have investigated a broadband multidimensional variation measurement of a force acting on a  mechanical oscillator taking into account asymmetry of interaction as well as optical loss. We found that the asymmetry does not prevent the complete removal of the quantum back action from the measurement result. 
The back action compensation is not complete in the presence of optical losses, it restricts the value of back action correlation in the two measurement channels  and, hence, sensitivity.

 \acknowledgments
The research of SPV has been supported by  Theoretical Physics and Mathematics Advancement Foundation “BASIS” (Contract No. 22-1-1-47-1), by the Interdisciplinary Scientific and Educational School of Moscow University ``Fundamental and Applied Space Research'' and by the TAPIR GIFT MSU Support of the California Institute of Technology. The reported here research performed by ABM was carried out at the Jet
Propulsion Laboratory, California Institute of Technology, under a contract with the National Aeronautics and Space Administration (80NM0018D0004).  This document has LIGO number P2400323.   

\appendix

.

\section{Standard Quantum Limit}\label{appSQL}

Here we recall details on the derivation of SQL for the force acting on the mechanical oscillator. Let us assume that we measure the phase quadrature of the resonant light reflected from the movable end mirror of a Fabry-Perot cavity. 

The single-sided power spectral density, corresponding to the unity signal to noise ratio, recalculated to the quadrature $f_{s\, a}$ of the normalized signal force (\ref{Fs}, \ref{fs}), acting during time $\tau$ is
\begin{align}
	\label{Sfa}
 S_{fa} &= 2\gamma_m \left(n_T+\frac 1 2\right)+ 
 	\frac{\gamma_m^2+\Omega^2}{\mathcal K} +\mathcal K\ge\\
 	&\ge  2\gamma_m \left(n_T+\frac 1 2\right)+ S_{SQL},\\
 	\label{SSQL}
 	& S_{SQL}= 	2\sqrt{\gamma_m^2 +\Omega^2}
\end{align}
Here parameter $\mathcal K$ is proportional to pump power, $\Omega$ is spectral frequency. Below we put angle $\psi_f=0$ in \eqref{Fs}. 

The approximate condition the force detection is
\begin{align}
 \label{approximate}
    f_{s\, a}&=\frac{F_{s0}}{\sqrt 2\sqrt{2\hslash m\omega_m}} \ge\sqrt{\int_0^{2\pi/\tau} S_f(\Omega)\,\frac{d\Omega}{2\pi} }=\\
    =&\sqrt{\left[2\gamma_m\left(n_T+\frac 1 2\right) +\frac{\gamma_m^2 + \frac{1}{3}\left[\frac{2\pi}{\tau}\right]^2}{\mathcal K} + \mathcal K\right] \frac{1}{\tau}} 
\end{align}
We should consider the case of short time $\tau$ 
\begin{equation}
 \label{condtau}
 \gamma_m\tau\ll 1   
\end{equation}
since in the opposite case of large $\tau$  the thermal limit restricts the sensitivity. Optimizing the measurement with respect of  $\mathcal K$ and taking into account of \eqref{condtau}, we obtain the minimum detectable force $F_{s0}$ amplitude 
\begin{align}
 \label{Fs0}
 \frac{F_{s0}^2}{4\hslash m \omega_m}&\ge  2\gamma_m\left(n_T+\frac 1 2\right)\,\frac{1}{\tau} +\frac{2}{\sqrt 3}\frac{2\pi}{\tau^2}
\end{align}
Here first term here describes the thermal limit whereas the second one introduces SQL:
\begin{align}
    F_{s0}^{SQL} =\frac{4}{\tau}\sqrt\frac{\pi \hslash m \omega_m}{\sqrt 3}
\end{align}
This formula is valid with an accuracy of a constant multiplier of about unity due to the approximation of the force envelope shape \eqref{approximate}.

We can use \eqref{SSQL} instead of \eqref{Sfa} and obtain 
\begin{align}
 \label{Fs02}
 \frac{\tilde F_{s0}^2}{4\hslash m \omega_m}&\ge  2\gamma_m\left(n_T+\frac 1 2\right)\frac{1}{\tau} +\frac{4\pi}{\tau^2}
\end{align}
The second terms in \eqref{Fs0} and in \eqref{Fs02} differ only by a multiplier of about unity. Therefore, we can use spectral density \eqref{SSQL} for SQL characterization in the frequency domain. 

\section{Derivation of the output quadrature} \label{appPSD}

We start from the equations \ref{quadAmp} for the amplitude quadratures
\begin{subequations}
	\begin{align}
		\label{ca+}   
		(\gamma_+ - i\Omega)c_{+a} &+  \eta_+ C_0  d_a = \sqrt {2 \gamma_{0+}}  { a}_{ +a} +\\
		& + \sqrt{2\gamma_{e+}} e_{+a},\nonumber\\
		\label{ca-} 
		(\gamma_- - i\Omega) c_{-a} &- \eta_- C_0  d_a = \sqrt {2 \gamma_-}  { a}_{-a}+\\
		& + \sqrt{2\gamma_{e-}} e_{-a},\nonumber\\
		\label{da}
		(\gamma_m - i\Omega)  d_a &-  C_0 \Big(\eta_+c_{+a}+ \eta_-c_{-a}\Big)=\\
		&= \sqrt {2 \gamma_m} q_a +  f_{s\,a},\nonumber
	\end{align}
\end{subequations}

From (\ref{ca+}, \ref{ca-}) we  find the back action term $\sim (\eta_+ c_{+a} + \eta_- c_{-a})$ and substitute it into \eqref{da} to derive
\begin{subequations}
\begin{align}
	\label{daAdd}
	\left(\Gamma_m  - i\Omega   \frac{}{}\right) & d_a
	= C_0 \frac{\eta_+(\sqrt {2 \gamma_+}a_{ +a}+\sqrt{2\gamma_e}e_{+a})}{\gamma_+ - i\Omega}  +\nonumber\\  
	+& C_0\frac{\eta_-(\sqrt {2 \gamma_-}a_{ -a}+\sqrt{2\gamma_e}e_{-a})}{\gamma_- - i\Omega}  \\
	& +\sqrt {2 \gamma_m} q_a +  f_{s\,a},\\
   \Gamma_m =\gamma_m + G, &\quad 	G  \equiv \left[\frac{\eta_+^2}{\gamma_+-i\Omega}- \frac{\eta_-^2}{\gamma_- -i\Omega}\right] C_0^2 
\end{align}
\end{subequations}
The optical damping rate $G$ appears due to non-symmetry of $\eta_\pm$ and $\gamma_\pm$.

Using the input-output relations
\begin{equation}
 \hat b_\pm= -\hat a_\pm + \sqrt{2\gamma_{0\pm}} \hat c_\pm
\end{equation}
We find the output amplitude quadratures and write them in form
\begin{subequations}
	\label{quadInDNS3}
	\begin{align}
		\label{ca+DDNS3}   
		b_{+a}  &= \xi_+\,  { a}_{ +a} + \mu_+ e_{+a} 
		-\frac{\sqrt{2\gamma_{0+}}\,\eta_+}{\gamma_+ -i\Omega}\, C_0  d_a,\\
		\label{ca-DDNS3} 
		b_{-a}  &= \xi_-\,  { a}_{ -a} + \mu_- e_{-a}
		+\frac{\sqrt{2\gamma_{0-}}\,\eta_-}{\gamma_- -i\Omega}\, C_0  d_a,\\
		\xi_\pm =&\frac{\gamma_{0\pm}-\gamma_{e\pm} + i\Omega}{\gamma_{0+} +\gamma_{e\pm} - i\Omega},\quad
		\mu_\pm=\frac{2\sqrt{\gamma_{0\pm}\gamma_{e\pm}} }{\gamma_{0\pm}+\gamma_{e\pm} -i\Omega}
	\end{align}
\end{subequations}

After substitution of \eqref{daAdd} into \eqref{quadInDNS3} we obtain
\begin{subequations}
 \begin{align}
   b_{+a}&=\frac{A_+}{\Gamma_m-i\Omega} 
   \label{b+Ma}
    \left\{\left(B_+-A_+\right)a_{+a} - A_-\, a_{-a} +\frac{}{}\right.\\  &+\sqrt{\frac{\gamma_{e+}}{\gamma_{0+}}}\left(B_{e+} - A_+\right) e_{+a} - \sqrt{\frac{\gamma_{e-}}{\gamma_{0-}}}A_- e_{-a}\\
   &\quad -\left.\left(\sqrt {2 \gamma_m} q_a +  f_{s\,a}\right)\right\} \\
   b_{-a}&=\frac{A_-}{\Gamma_m-i\Omega} 
   \label{b-Ma}
   \left\{\left(B_- +A_-\right)a_{-a} + A_+\, a_{+a} +\frac{}{}\right.\\  &+\sqrt{\frac{\gamma_{e-}}{\gamma_{0-}}}\left(B_{e-} + A_-\right) e_{-a} + \sqrt{\frac{\gamma_{e+}}{\gamma_{0+}}}A_+ e_{+a}\\
   &\quad +\left.\left(\sqrt {2 \gamma_m} q_a +  f_{s\,a}\right)\right\} \\
   B_\pm &=\frac{(\gamma_{0\pm} -\gamma_{e\pm} +i\Omega)(\Gamma_m-i\Omega)}{\sqrt{2\gamma_{0\pm}}\,\eta_\pm C_0}, \\  
   A_\pm &= \frac{\sqrt{2\gamma_{0\pm}}\,\eta_\pm C_0}{(\gamma_\pm -i\Omega)}, \ B_{e\pm}=\frac{\sqrt{2\gamma_{0\pm}}(\Gamma_m-i\Omega)}{\eta_\pm C_0}.
\end{align}
\end{subequations}

We measure $b_{+a}$ and $b_{-a}$ separately and take its weighted sum to perform the measurement. Let consider $\Sigma$ to be a measured linear combination ofthe amplitude quadratures
\begin{align}
  \label{Sigma1}
  \Sigma = z_+ \, \frac{b_{+a}\big(\Gamma_m -i\Omega\big)}{A_+\big(z_- -z_+\big)}+ z_- \, \frac{b_{-a}\big(\Gamma_m -i\Omega\big)}{A_-\big(z_- -z_+\big)}
\end{align}
where $z_\pm (\Omega)$ are complex functions depending on spectral frequency $\Omega$, which should be chosen for post-processing in an optimal way. The parameter $\Sigma$ is chosen so that the normalized signal term $\big( f_{s\,a}\big)$ has coefficient of unity. 
\begin{subequations}
	\begin{align}
	\begin{split}
\Sigma &= B_+\left[y-\frac 1 2 + Y_+\right] a_{+a}
		+ B_-\left[y+\frac 1 2 + Y_-\right]a_{-a} + \\
		&+ \sqrt\frac{\gamma_{e+}}{\gamma_{0+}} B_{e+}\left[y-\frac 1 2 + Y_{e+}\right] e_{+a}+\\
		+& \sqrt\frac{\gamma_{e-}}{\gamma_{0-}} B_{e-} \left[y+\frac 1 2 + Y_{e-}\right] e_{-a}+ \sqrt {2 \gamma_m} q_a +  f_{s\,a},
\end{split} \label{Sigma}
		\\
		& y+\frac 1 2= \frac{Z_-}{Z_--Z_+} ,\quad -\frac 1 2 +y =\frac{Z_+}{Z_--Z_+} \\ 
		\label{Ypm}
		&Y_\pm=\frac{A_\pm}{B_\pm} ,\quad Y_{e\pm}= \frac{A_\pm}{B_{e\pm}}
	\end{align}
\end{subequations}
Here we introduced one optimization parameter $y$ instead of two parameters $z_\pm$.


\end{document}